\documentclass[a4paper]{article}
\usepackage{RRA4}
\RRNo{6489}
\usepackage{hyperref}
\usepackage{amsmath}
\usepackage{graphicx}
\usepackage{amssymb}
\usepackage{eqnarray}
\usepackage{theorem}                    % jolis théorèmes 
\usepackage[lined, algoruled]{algorithm2e}
\usepackage[latin1]{inputenc}
\usepackage[T1]{fontenc}
\usepackage{subfigure}

\DeclareMathOperator{\conf}{conf}
\DeclareMathOperator{\scg}{scg}

\DeclareMathOperator{\makesw}{makesw}

\RRdate{Avril 2008}
\RRauthor{% les auteurs
Bruno Gaume\thanks{IRIT--UPS, Toulouse F-31062 Cedex 4, France}%
  \and
Fabien Mathieu\thanks{Orange Labs, 38--40 rue du Général Leclerc, 92794 Issy-les-Moulineaux Cedex}%
}
\authorhead{Bruno Gaume \& Fabien Mathieu}
\RRtitle{Petit-mondisation par marches aléatoires}
\RRetitle{From Random Graph to Small World by Wandering}
\titlehead{From Random Graph to Small World by Wandering}
\RRresume{De nombreuses études montrent un fait remarquable qui est que la plupart des réseaux dits de terrain possèdent des propriétés identiques bien particulières et font partie de la classe des graphes \emph{petit-monde}. Un autre fait tout aussi remarquable est que cette classe des petits mondes est très petite au regard de l'ensemble des graphes possibles.
Dans cet article, nous proposons une méthode de production de graphes petit-monde au moyen de marches aléatoires.
}
\RRabstract{Numerous studies show that most known \emph{real-world} complex networks share similar properties in their
connectivity and degree distribution.
They are called \emph{small worlds}.
This article gives a method to turn random graphs into Small World graphs by the dint of random walks.}
\RRmotcle{Graphes aléatoires, petits mondes, marches aléatoires}
\RRkeyword{Random graphs, small worlds, random walks}
\RRprojet{GANG}
\RRtheme{\THCom}% \THCog \THSym \THNum \THBio} % cas de 5 themes
\URRocq % pour ceux qui sont au centre de la France
\begin{document}
\makeRR   % cas d'un rapport de recherche

\newcommand{\mG}{\mathcal{G}}
\newcommand{\Ned}{\mathbb{N}}
\newcommand{\Red}{\mathbb{R}}
\newcommand{\dist}{\mathop{\mathrm{dist}}}
\newcommand{\abs}[1]{\left\vert#1\right\vert}
\newcommand{\eq}[1]{\begin{equation} #1
\end{equation}}

\section{Introduction}
\label{sec:intro}

In 1998, Watts and Strogatz showed that many real graphs, coming from different fields, share similar properties~\cite{watts98collective}. This has been confirmed by many studies since this seminal work~\cite{albert99diameter,kumar00web,broder00graph,adamic00power,faloutsos99powerlaw,govindan00heuristics,barabasi1999emergence,newman01scientific,barabasi01evolution,montoya00small,ferrer01small,albert02statistical}. The concerned fields include, but are not limited to:  epidemiology (contact graphs, \ldots), economy (exchange graphs, \ldots), sociology (knowledge graphs,\ldots), linguistic (lexical networks, \ldots), psychology (semantic association graphs,\ldots), biology (neural networks, proteinic interactions graphs), IT (Internet, Web)\ldots We call such graphs real-world complex networks, or small-world networks.

The common properties of real-world complex networks are a low diameter, a globally sparse but locally heavy edge density, and a heavy-tailed degree distribution. The combination of these property is very unlikely in random graphs, explaining the interest that those networks have arisen in different scientific communities.

In this article, we propose a method to generate a graph with small-world properties from random graph. This method, which is based on random walks, may be a first step in order to understand why graphs from various origins share a common structure.

In Section \ref{sec:sw_structure}, we briefly state the properties used to decide wheter a given graph is small world or not. In Section \ref{sec:state_art}, we survey the different existing methods to generate complex networks. In Section \ref{sec:Confl_by_random_walk}, we analyse the dynamics or random walks in a graph, and in Section \ref{sec:random_to_small} we propose a new method to construct small worlds by \emph{wandering} on random graphs. Section \ref{sec:conclusion} concludes.

\section{Small Worlds Structure}
\label{sec:sw_structure}

let $G=(V,E)$ be a reflexive, symmetric graph with $n=\abs{V}$ nodes and $m=\abs{E}$ edges. $G$ is called \emph{small world} if the following properties are verified:
\begin{description}
\item[Edge sparsity] Small world graphs are sparse in edges, and the average degree stay low: $m=O(n)$ or $m=O(n\log(n))$
\item[Short paths] The average path length (denoted $\ell$) is close to the average path length $\ell_\text{rand}$ in the main connected component of $G(n,m)=\mG(n,\frac{m-n}{n(n-1)})$ Erdös-Rényi graphs. According to~\cite{erdos59random}, for $d:=\frac{m-n}{n}\geq (1+\epsilon)\log(n)$, $\mG(n,\frac{m-n}{n(n-1)})$ is almost surely connected, and $\ell_\text{rand}\approx \frac{\log(n)}{\log(d)}$. ($l=O(\log(n))$).
\item[High clustering] The clustering coefficient, $C$, that expresses the probability that two disctinct nodes adjacent to a given third node are adjacent, is an order of magnitude higher than for Erdös-Rényi graphs: $C>>C_\text{rand}=p=\frac{m-n}{n(n-1)}$. This indicates that the graph is locally dense, although it is globally sparse.
\item [Heavy-tailed degree distribution]
\end{description}

\begin{figure}[ht]
	\centering
		\subfigure[linear scales]{\includegraphics[width=.4\textwidth]{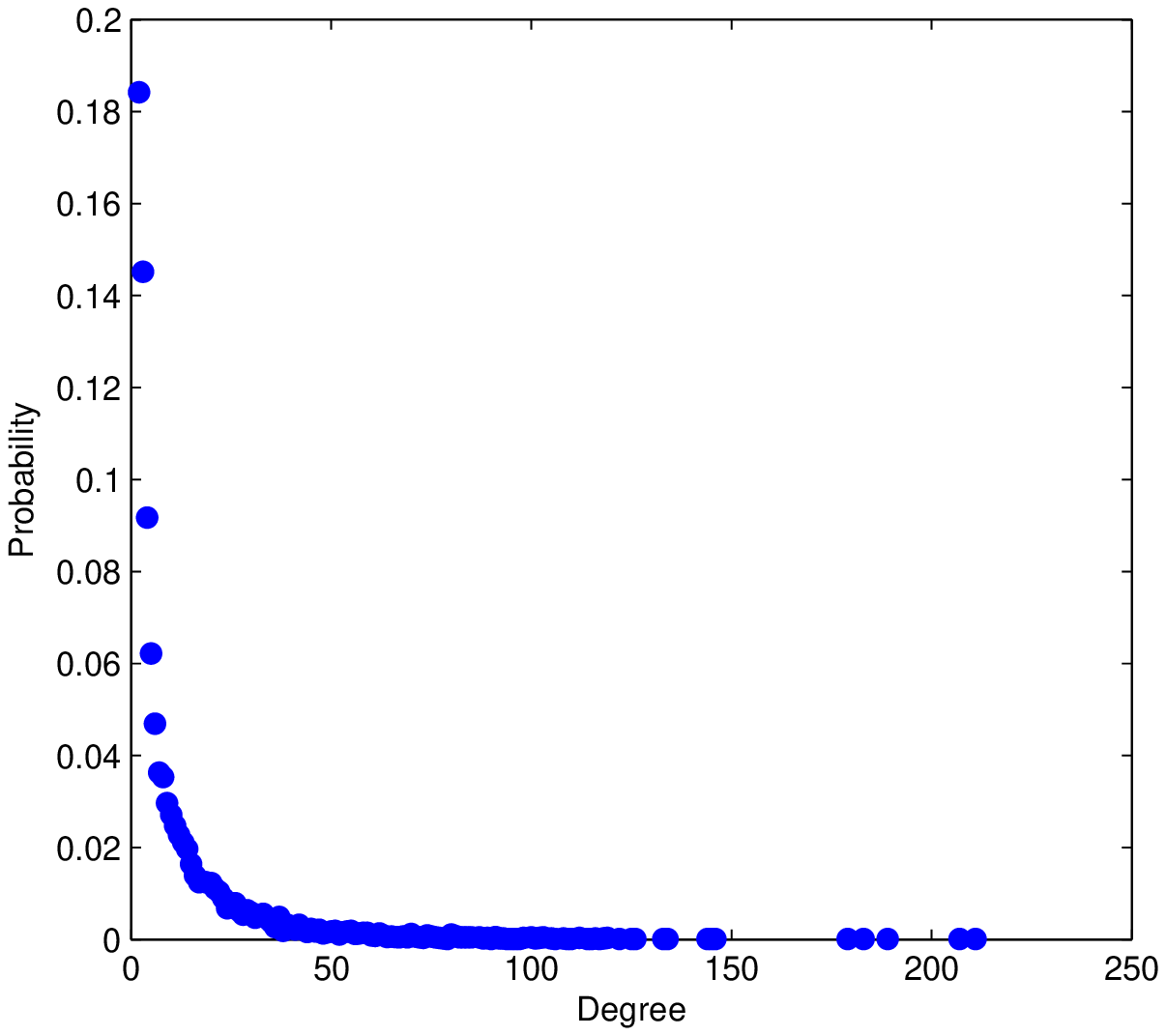}}
		\subfigure[logarithmic scales]{\includegraphics[width=.4\textwidth]{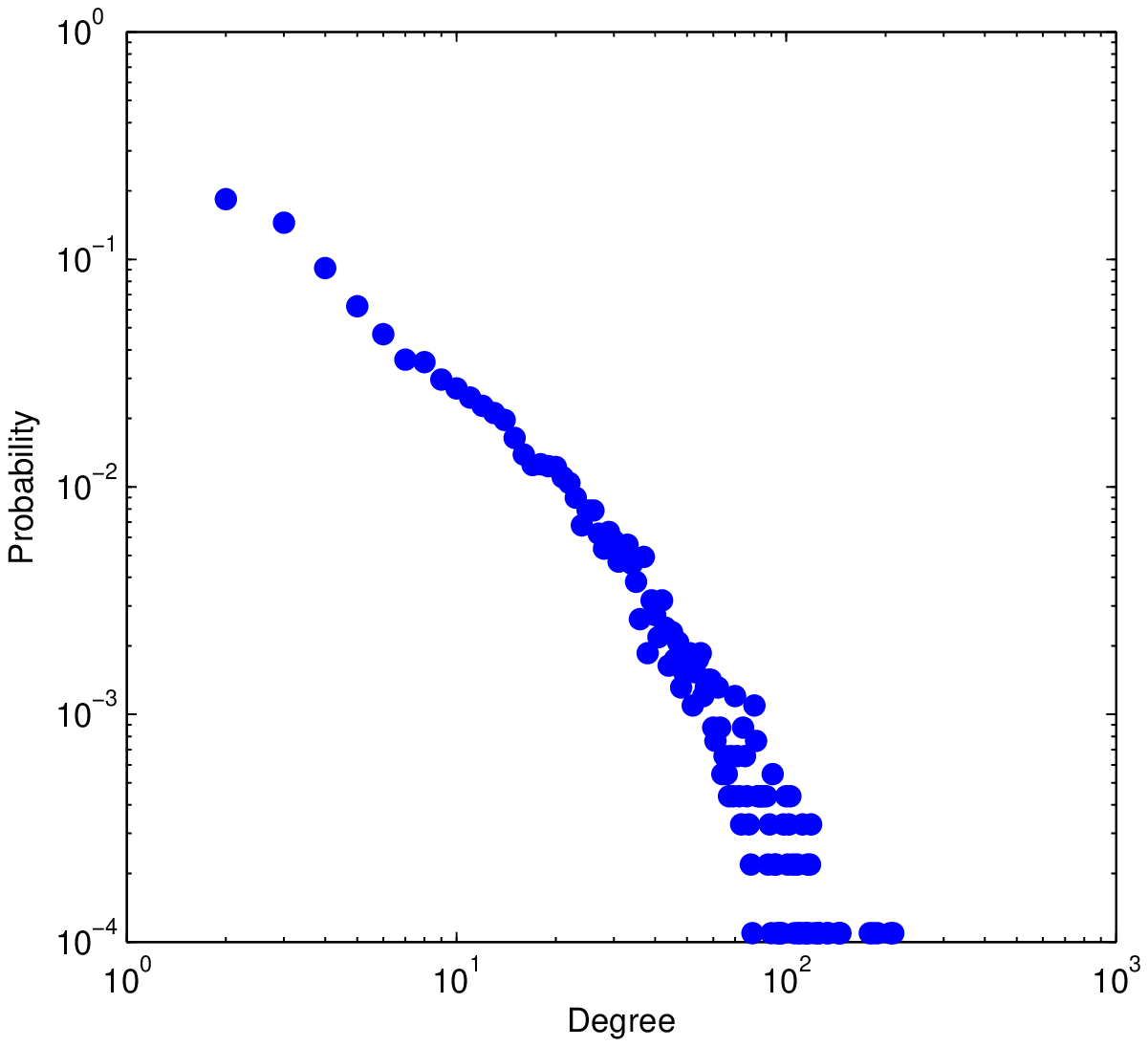}}
	\caption{Degree distribution of $G_c$}
	\label{fig:dicosyn_loglog}
\end{figure}

\paragraph{Example:} DicoSyn.Verbe\footnote{DicoSyn is a french synonyms dictionnary built from seven canonical french dictionnaries (Bailly, Benac, Du Chazaud, Guizot, Lafaye, Larousse et Robert). The ATILF (\url{http://www.atilf.fr/}) extracted the synonyms, and the CRISCO (\url{http://elsap1.unicaen.fr/}) consolidated the results. DicoSyn.Verbe is the subgraph induced by the verbs of Dicosyn: an edge exists between two verbs $a$ and $b$ iff DicoSyn tells $a$ and $b$ are synonyms. Therefore DicoSyn.verbe is a symmetric graph, made reflexive for convenience. A visual representation based on random walks~\cite{gaume04balades} can be consulted on \url{http://Prox.irit.fr}.} is a reflexive symmetric graph with $9043$ nodes and $110939$ edges. For sake of convenience, we only consider the main connected component $G_c$ of DicoSyn, which admits $8835$ nodes and $110533$ edges. With an average degree of $12.5$, $G_c$ is sparse. Other parameters of $G_c$ are $\ell\approx 4.17$ (to compare with $\ell_\text{rand}=3.71$) and $C\approx 0.39$ (to compare with $C_\text{rand}=p=0.0013$). The degree distribution is heavy-tailed, as shown by Figure~\ref{fig:dicosyn_loglog} (a least-square method gives a slope of $-2.01$ with a confidence $0.96$). Therefore $G_c$ verifies the four properties of a small world.

\begin{figure}[ht]
	\centering
		\subfigure[linear scales]{\includegraphics[width=.4\textwidth]{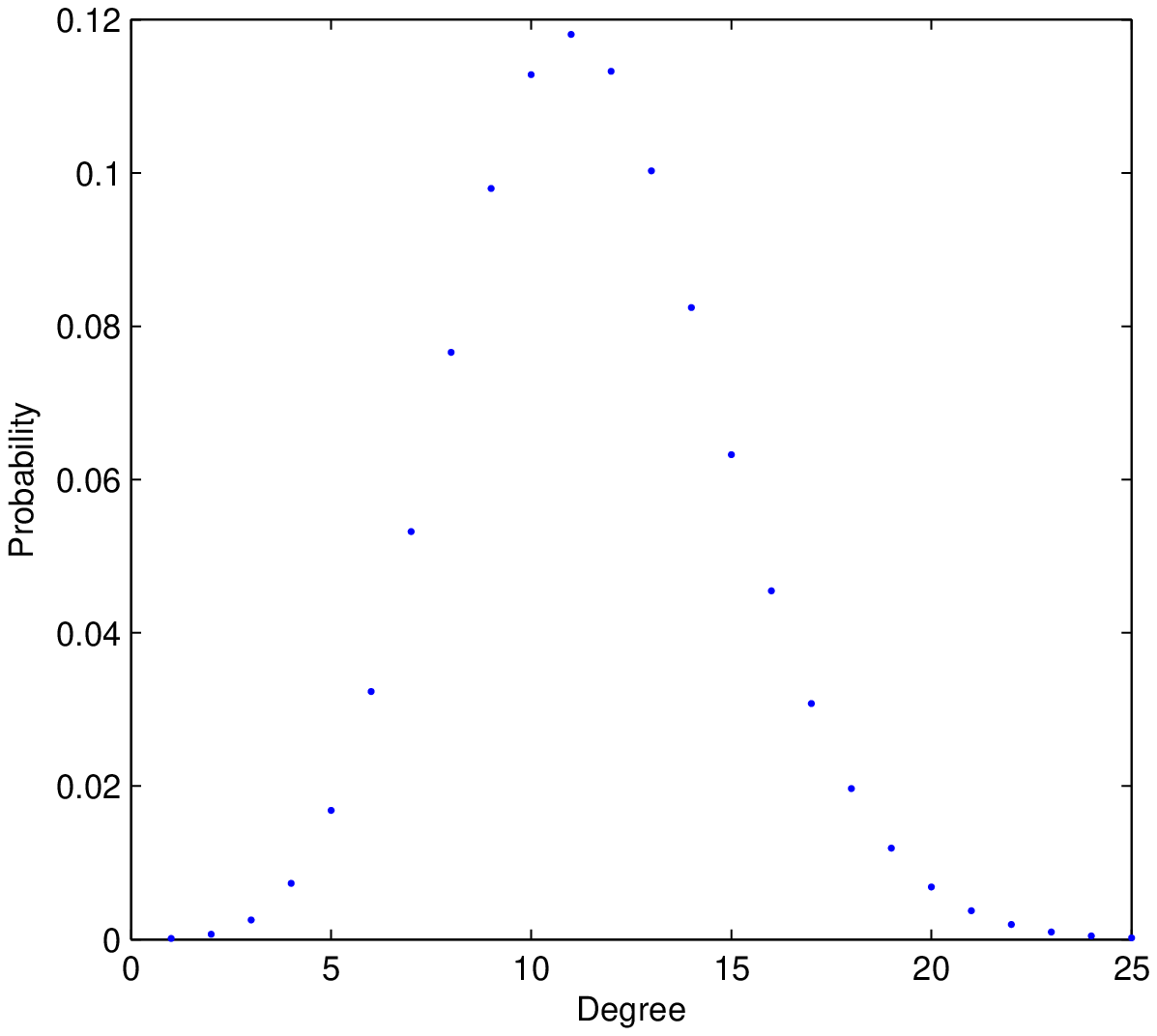}}
		\subfigure[logarithmic scales]{\includegraphics[width=.4\textwidth]{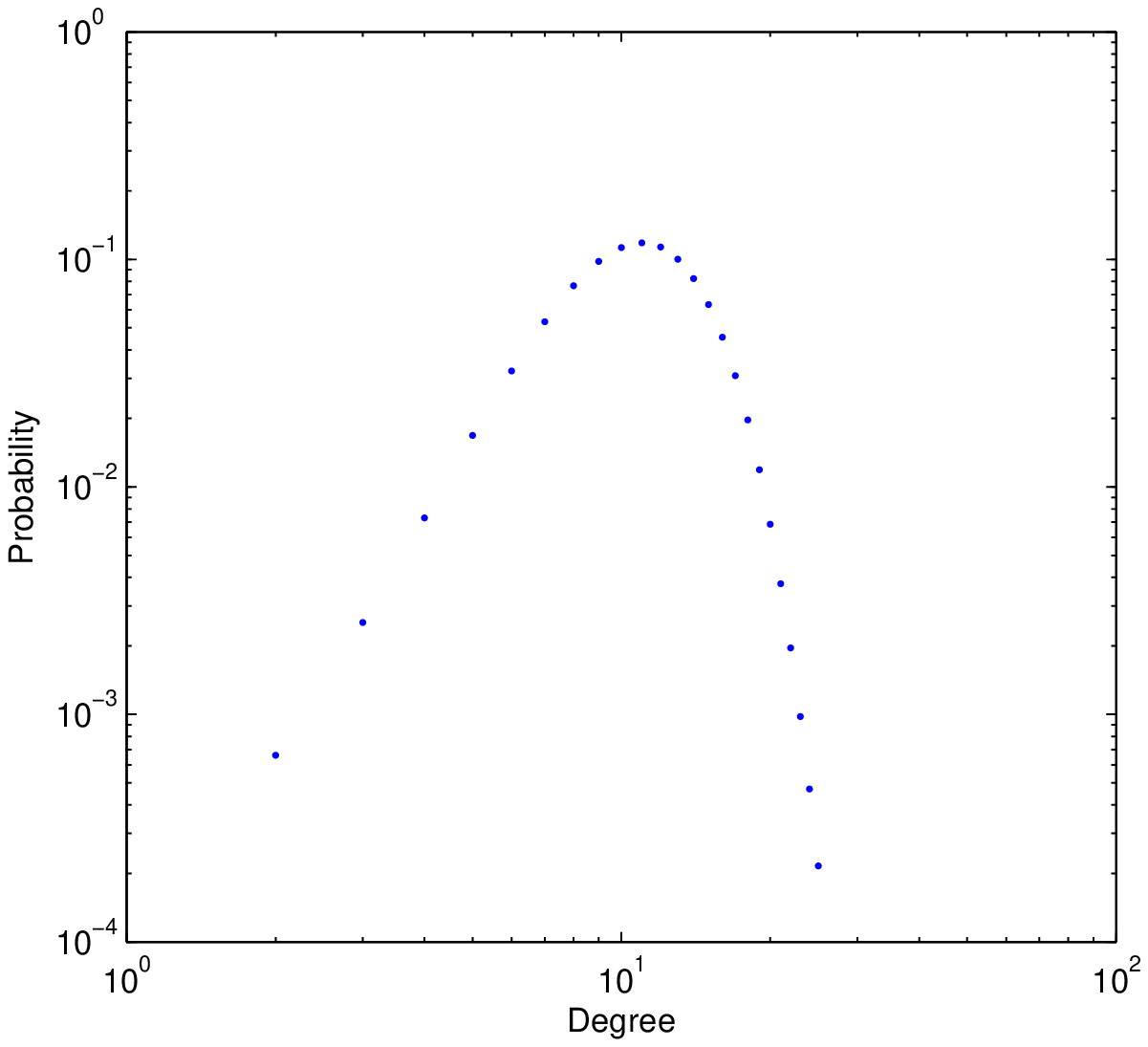}}
	\caption{Degree distribution of a typical $\mG(n,p)$ graph}
	\label{fig:erdos_distrib}
\end{figure}

Note, that the degree distribution for random Erdös-Rényi graphs is far from being heavy-tailed. It is in fact a kind of Poisson distribution : the probability that a node of a $\mG(n,p)$ graph has degree $k$ is $p(k)=p^k(1-p)^{n-1-k}{n-1 \choose k}$.  Figure~\ref{fig:erdos_distrib}, where the degree distribution of a Erdös-Rényi graph with same number of nodes and average degree than $G_c$ is plotted. This illustrates how a small world compares to a $\mG$ graph with same number of nodes and expected degree:
\begin{itemize}
	\item Same sparsity (by construction),
	\item Similar average path length,
	\item Higher clustering,
	\item Heavy-tailed distribution (instead of Poisson distribution)
\end{itemize}

In~\cite{albert02statistical}, Albert and Barabasi have made a survey on existing complex networks studies, including~\cite{albert99diameter,kumar00web,broder00graph,adamic00power,faloutsos99powerlaw,govindan00heuristics,barabasi1999emergence,newman01scientific,barabasi01evolution,montoya00small,ferrer01small,watts98collective}. Some of their findings are presented in Table~\ref{tab:realworld} along $G_c$'s properties.

\begin{table}[ht]
	\centering
		\begin{tabular}{|l|r|r|r|r|r|r|}
			\hline
			Name & $n$ & $<k>$ & $\ell$ & $C$ & $\gamma$ & $r_2$ \\
			\hline
			DicoSyn.Verbes & $8835$ & $11.51$ &  $4.17$ & $0.39$ & $2.01$ & $0.96$ \\
			\hline
			Internet routers & $150000$ & $2.66$ & $11$ &  & $2.4$ & \\
			\hline
			Movie actors & $212250$ & $28.78$ & $4.54$ & $0.79$ &  $2.3$ &  \\
			\hline
			Co-authorship, SPIRES & $56627$ & $173$ & $4.0$ & $0.726$ & $1.2$ & \\
			\hline
			Co-authorship, math. & $70975$ & $3.9$ & $9.5$ & $0.59$ & $2.5$ &\\
			\hline
			Co-authorship, neuro. & $209293$ & $11.5$ & $6$ & $0.76$ & $2.1$ & \\
			\hline
			Ythan estuary food web & $134$ & $8.7$ & $2.43$ & $0.22$ & $1.05$ &\\
			\hline
			Silwood Park food web & $154$ & $4.75$ & $3.40$ & $0.15$ & $1.13$ &\\
			\hline
			Words, synonyms & $22311$ & $13.48$ & $4.5$ & $0.7$ & $2.8$ &\\
			\hline
		\end{tabular}
	\caption{Main properties of some complex networks}
	\label{tab:realworld}
\end{table}

\section{Generating Small Worlds: State of Art}
\label{sec:state_art}

Small-world networks have been studied intensely since they were first described in Watts and Strogatz~\cite{watts98collective}. Researchs have been done in order to be able to generate random datasets with well-known characteristics shared by social networks. Most papers focus on either the clustering and diameter, or on the power-law.

\subsection{Clustering and diameter property}

Watts and Strogatz \cite{watts98collective}, and Kleinberg \cite{kleinberg00small} have studied families of random graphs that share the clustering and diameter properties of small worlds. Watts and Strogatz model consist in altering a regular ring lattice by rewiring randomly some links. In Kleinberg's model, a $d$-dimensional grid is extended by adding extra-links of which the range follows a $d$-harmonic distribution.

Note, that both models fail to capture the heavy-tail property met in real complex networks (they are almost regular).

\subsection{Heavy-tail property}

There is a lot of research devoted on the production of random graphs that follow a given degree distribution~\cite{bollobas01random,luczak92sparse,molloy95critical,newman02assortative}. Such generic models easily produce heavy-tailed random graphs if we give them a power law distribution.

On the field of specific heavy-tailed models, there is Albert and Barabasi preferential attachment's model~\cite{albert02statistical,barabasi1999emergence}, in which links are added one by one, and where the probability that an existing node receives a new link is proportional to its degree. A more flexible version of the preferential attachment's model is the fitness model~\cite{adamic00power,bianconi01bose}, where a pre-determined fitness value is used in the process of link creation. 

Lastly, Aiello \emph{et al.} proposed a model called \emph{$\alpha,\beta$ graphs}~\cite{aiello00random}, that encompasses the class of power law graphs.

\subsection{Others models}

Other models of graph generation are Guillaume and Latapy's \emph{All Shortest Paths}~\cite{guillaume05complex}, where one construct a graph by extracting the shortest paths of a random graph, and the Dorogovtsev-Mendes model~\cite{dorogo02evolution}. Note, that the latter captures all desired properties, but is not realistic.

\section{Confluence \& Random Walk in Networks}
\label{sec:Confl_by_random_walk}
\subsection{Random Walk in Networks}
\label{subsec:random_walk}

Just like Section~\ref{sec:sw_structure}, $G=(V,E)$ is a reflexive, symmetric graph with $n=|V|$ nodes and $m=|E|$ edges. We assume that a particle wanders randomly on the graph:
\begin{itemize}
	\item At any time $t\in\Ned$ the particle is on a node $u(t)\in V$;
	\item At time $t+1$, the particle reaches a uniformly randomly selected neighbor of $u(t)$.
\end{itemize}

This process is an homogeneous Markov chain for on $V$. A classical way to represent this chain is a $n\times n$ stochastic matrix $[G]$: \\
\eq{[G]=(g_{u,v})_{u,v\in V}\text{, with } g_{u,v} = \left\{ \begin{gathered}\frac{1}{\deg(u)}\text{ if }u \to v\text{,}\hfill \\
0\text{ else.}\hfill \\
\end{gathered}\right.\label{eq:def-A}}

Because $G$ is reflexive, no node has null degree, so the underlying Markov chain $[G]$ is well defined.
For any initial probability distribution $P_0$ on $V$ and any given integer $t$, $P_0[G]^t$ is the result of the random walk of length $t$ starting from $P_0$ whose transitions are defined by $[G]$.
More precisely, for any $u$, $v$ in $V$, the probability $P_t$ of being in $v$ after a random walk of length $t$ starting from $u$ is equal to $(\delta_u[G]^t)_v=([G]^{t})_{u,v}$, where $\delta_u$ is the certitude of being in $u$.
One can demonstrate, by the dint of the Perron-Frobenius theorem \cite{ne94perronfrobenius}, that if $G = (V,E)$ is a connected, reflexive and symmetric graph, then:
\begin{equation}
	\forall u, v \in V, \lim_{t\rightarrow \infty} (\delta_u[G]^t)_v = \lim_{t\rightarrow \infty}([G]^{t})_{u,v}= \frac{\deg(v)}{\sum_{x\in V} \deg(x)}
	\label{eq:lim-infini}
\end{equation}
In other words, given than $t$ is large enough, the probability of being on node $v$ at time $t$  is proportional to the degree of $V$, and no longer depends on the departure node $u$.
\subsection{Confluence in Networks}
\label{subsec:confluence}

Equation~\eqref{eq:lim-infini} tells that the only information retained after an infinite random walk is the degree of the nodes. However, some information can be extracted from transitional states. For instance, assume the existence of three nodes $u$, $v_1$ and $v_2$ such that
\begin{itemize}
	\item $u$, $v_1$ and $v_2$ belong to the same connected component,
	\item $v_1$ is \emph{close} from $u$, in the sense that many short paths exist between $u$ and $v_1$,
	\item $v_2$ is \emph{distant} from $u$,
	\item $v_1$ and $v_2$ have the same degree.
\end{itemize}

From \eqref{eq:lim-infini}, we know that the sequences $(([G]^t)_{u,v_{1}})_{1\leq t}$ and $(([G]^t)_{u,v_{2}})_{1\leq t}$ share the same limit, that is $\deg(v_{1})/\sum_{x\in V} \deg(x)=\deg(v_{2})/\sum_{x\in V} \deg(x)$.

However these two sequences are not identical. Starting from $u$, the dynamic of the particle's trajectory on its random walk is completely determined by the graph's topological structure, and after a limited amount of steps $t$, one should expect a greater value for $(([G]^t)_{u,v_{1}})$ than for $(([G]^t)_{u,v_{2}})$ because $v_1$ is closer from $u$ than $v_2$.

This can be verified on the graph of french verbs $\mG_c$, with:
\begin{itemize}
	\item $u=\text{\emph{déshabiller}}$ (``to undress''),
	\item $v_1=\text{\emph{effeuiller}}$ (``to thin out''),
	\item $v_2=\text{\emph{rugir}}$ (``to roar''),
\end{itemize}

Intuitively, \emph{effeuiller} should be closer (in $G_c$) to \emph{déshabiller} than \emph{rugir}, because this is the case semantically. Also \emph{effeuiller} and \emph{rugir} have the same degree ($11$).

\begin{figure}[ht]
	\begin{center}
		\subfigure[French verbs graph $G_c$]{\includegraphics[width=.45\textwidth]{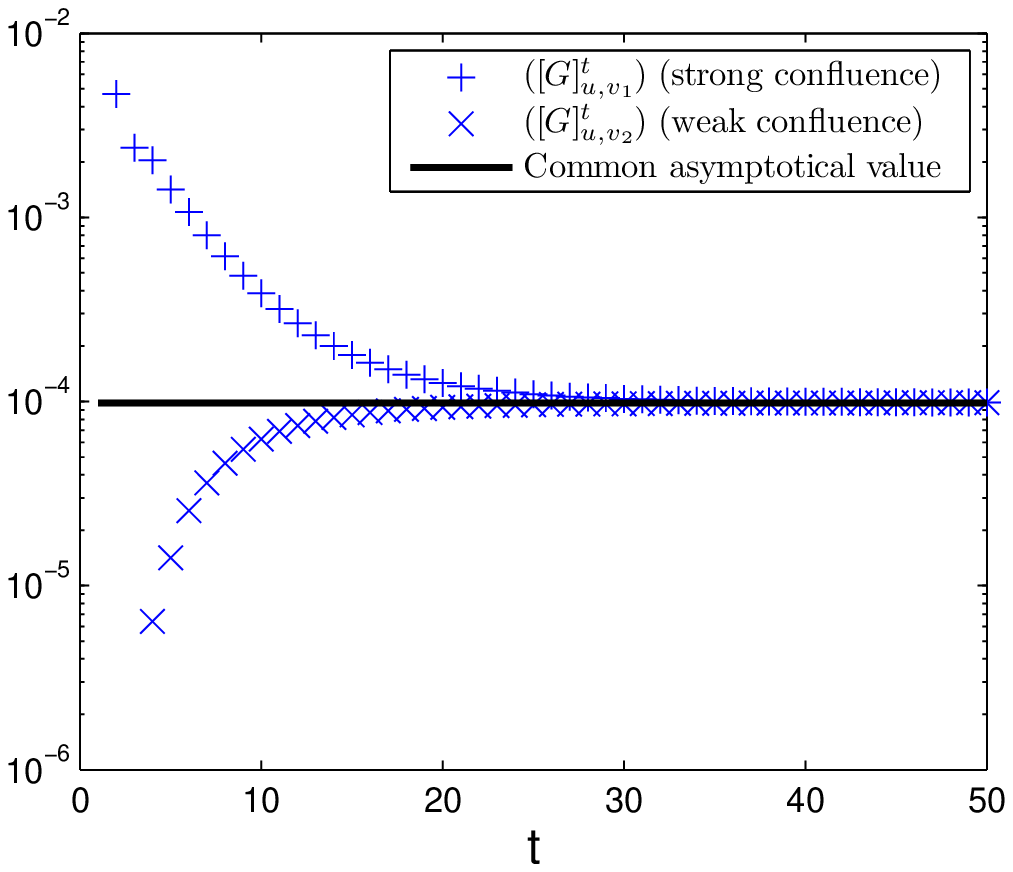} \label{fig:conflg1}}
		\subfigure[Random graph]{\includegraphics[width=.45\textwidth]{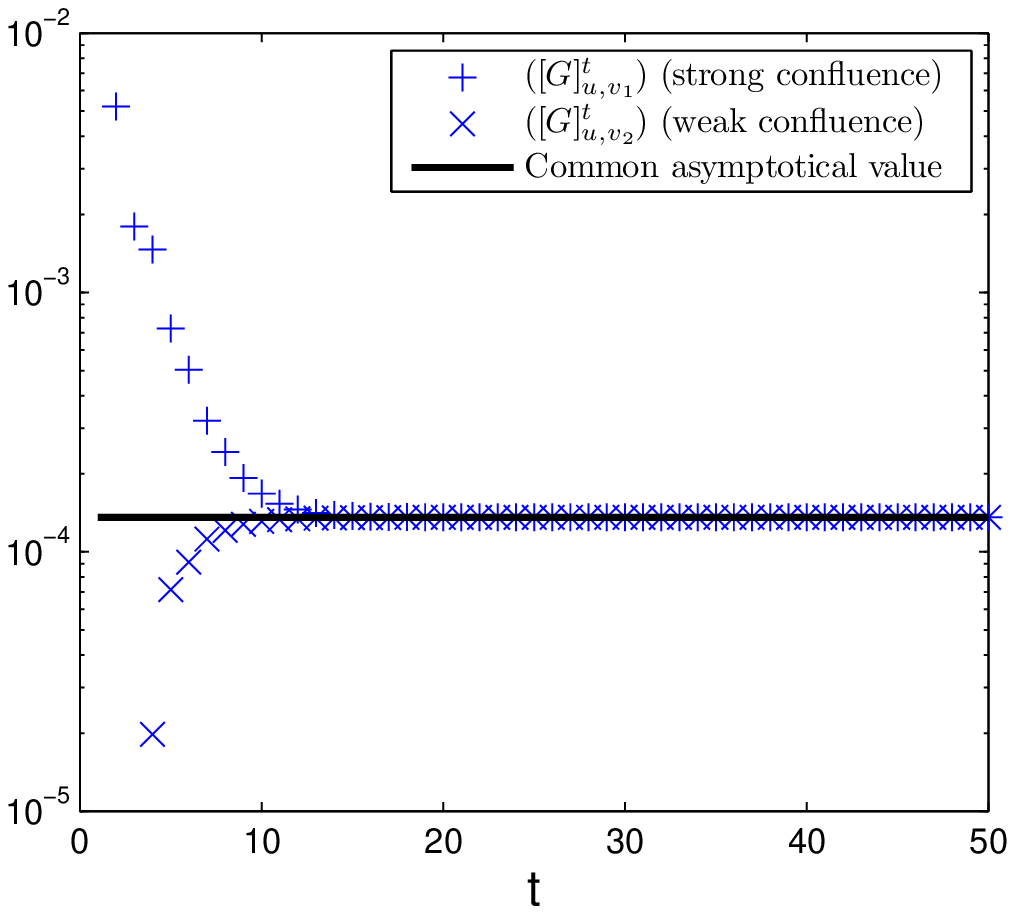} \label{fig:conflgrandG13}}
	\end{center}
	\caption{$(([G]^t)_{u,v_{1}})$ and $(([G]^t)_{u,v_{1}})$ for $G_c$ and a random graph}
	\label{fig:confluences}
\end{figure}

The values of $(([G]^t)_{u,v_{1}})$ and $(([G]^t)_{u,v_{2}})$ with respect to $t$ are shown in Figure~\ref{fig:conflg1}, along with the common asymptotic value $\frac{11}{\sum_{x\in V} \deg(x)}$.

One can observe that, after a few steps, $(([G]^t)_{u,v_{1}})$ is above the asymptotic value. We claim that this is typical of nodes that are close to each other, and call this phenomenum \emph{strong confluence}. On the other hand, $(([G]^t)_{u,v_{2}})$ is always below the asymptotic value (\emph{weak confluence}).

One could think that the existence of strong and weak confluences is typical of graphs with high clustering, because the notion of \emph{closeness} sounds like belonging to a same community. However, strong and weak confluences also occur in graphs with low clustering coefficients, such as Erdös-Rényi random graphs. For example, Figure~\ref{fig:conflgrandG13} shows $(([\mG]^t)_{u,v_{1}})$ and $(([\mG]^t)_{u,v_{2}})$ for three nodes $u$, $v_1$ and $v_2$ carefully selected in $\mG$ an Erdös-Rényi graph with same number of nodes and average degree than $G_c$.

Figure~\ref{fig:conflgrandG13} is very similar to Figure~\ref{fig:conflg1}. This points out that the concept of confluence exists in random graphs like it does in small worlds. In the following Section, we will use this to turn random graphs into small-worlds.

\section{From Random Graph to Small World by Wandering}
\label{sec:random_to_small}

Now we want to use the concept of confluence to provide a way to construct small-world like graphs. In order to do that we introduce the mutual confluence $\conf$ between two nodes of a graph $G$ at a time $t$:

\begin{equation}
	\conf_G(u,v,t)= \max([G]^t_{u,v}, [G]^t_{v,u})
	\label{eq:c_def}
\end{equation}

For \emph{not too large} values of $t$, a strong mutual confluence between two nodes may indicate that those nodes are close. We claim that a good way to obtain a small world from a random graph is to set edges between the pairs of nodes with the highest confluence.

\subsection{Extracting the confluence graph}

Given an input graph $G_{in}=(V,E_{in})$, symmetric and reflexive,  with $n$ nodes and $m_{in}$ edges, a time parameter $t$ and a target number of edges $m$, one can extract a strong confluence graph $G=\scg(G_{in},t,m)$ defined by:

\begin{itemize}
	\item $G$ a symmetric, reflexive graph with the same nodes than $G_{in}$ and $m$ edges,
	\item $\forall r\neq s,u\neq v\in V$, if $(r,s)\in E$ and $(u,v)\notin E$, then
		 $\conf_{G_{in}}(r,s,t) \geq \conf_{G_{in}}(u,v,t)$.
\end{itemize}
\begin{algorithm}
\dontprintsemicolon
\caption{$\scg$ (strong confluence graph), extract highest confluences}
\label{algo:r2sw}
\SetLine
\KwIn{An undirected graph $G_{in}=(V,E_{in})$, with $n$ nodes and $m_{in}$ edges\;
A walk length $t\in\Ned^*$\;
A target number of edges $m\in[n,n^2]$\;}
\KwOut{A graph $G=(V,E)$, with $n$ nodes and $m$ edges}
\Begin{
$E \longleftarrow \emptyset$\;
\For{$i\leftarrow 1$ \KwTo $n$}{
$E \longleftarrow E\cup \{(i,i)\}$ \tcc*[f]{Make $G$ reflexive}} \;
$M\longleftarrow n$\;
\While(\tcc*[f]{Is there unset edges?}){$M< m$}{
\lnlset{ligne:a}{(a)}$(r,s)\longleftarrow\arg\max_{(u,v)\notin E}([G_{in}]^t_{u,v})$\; 
\lnlset{ligne:b}{(b)}$E \longleftarrow E\cup \{(r,s)\}$ \; 
\lnlset{ligne:c}{(c)}$E \longleftarrow E\cup \{(s,r)\}$ \tcc*[f]{Stay symmetric}\;
$M\longleftarrow M+2$\;
}
}
\end{algorithm}
Algorithm~\ref{algo:r2sw} proposes a way to construct $\scg(G,t,m)$. Note, that because of possible confluences with same values, line~\ref{ligne:a} is not deterministic. Furthermore, there is no guarantee that the strong confluence graph is unique, but the possible graphs can only differ by their (few) edges of lowest confluence. In practice, confluences are distinct most of the time\footnote{If uniqueness really matters, it suffices to use a total order on the pairs of $V$ in order to break ties in line~\ref{ligne:a}.}

\subsection{Making Small-Worlds}

\begin{algorithm}
\dontprintsemicolon
\caption{$\makesw$, Making a small world}
\label{algo:makesw}
\SetLine
\KwIn{A target number of nodes for the output graph $n\in\Ned$\;
A target number of edges for the random graph $m_{in}\in\Ned$\;
A walk length $t\in\Ned^*$\;
A target number of edges $m\in\Ned$\;}
\KwOut{A graph $G=(V,E)$, with $n$ nodes and $m$ edges}
\Begin{
$G_{in} \longleftarrow$ a symmetric, reflexive, Erdös-Rényi Random Graph with $n$ nodes and $m_{in}$ edges\;
$G \longleftarrow \scg(G_{in},t,m)$\;
$G \longleftarrow$ largest connected component of $G$\;
}
\end{algorithm}

We propose to construct graphs with a small-world structure by extracting the confluences of  Erdös-Rényi graphs, as described in Algorithm~\ref{algo:makesw}. Note, that the confluence extraction may produce disconnected graphs. Therefore we have to select the main connected component if we want to study properties like diameter. However, our experiments show that the size of the main connected component is always more than $80\%$, so this is not such a big issue.

\subsection{Validation}

In order to obtain good small-worlds, the values $n$, $m_{in}$, $m$ and $t$ must be carefully selected. In the following, we set $n=1000$, $m_{in}=4000$, and $m=10000$, and we focus on the importance of the parameter $t$.

Like stated in Section~\ref{sec:sw_structure}, there is no strict definition of a small-world, but typical values for diameter, clustering and degree distribution. We arbitrary propose to say that $G=\makesw(n,m_{in},t,m)$ is small-world shaped if it verifies:
\begin{itemize}
	\item $m\leq 10n\log(n)$ (verified for $n=1000$, $m=10000$),
	\item its clustering coefficient $C_G$ is greater than $\frac{10m}{n^2}$,
	\item its diameter is lower than $3\log(n)$,
	\item a least square fitting on the degree log-log distribution gives a negative slope of absolute value $\lambda$ greater than $1$, with a correlation coefficient $r^2$ grater than $0.8$.
\end{itemize}
\paragraph*{Remark}The power law estimation we give is not very accurate (see for instance~\cite{newman04power}). However, giving a correct estimation of the odds that a given discrete distribution is heavy-tailed is a difficult issue (\cite{goldstein04problems,clauset07power}), and refining the power-law estimation is beyond the scope of this paper.

\begin{figure}[ht]
	\begin{center}
\includegraphics[width=1\textwidth]{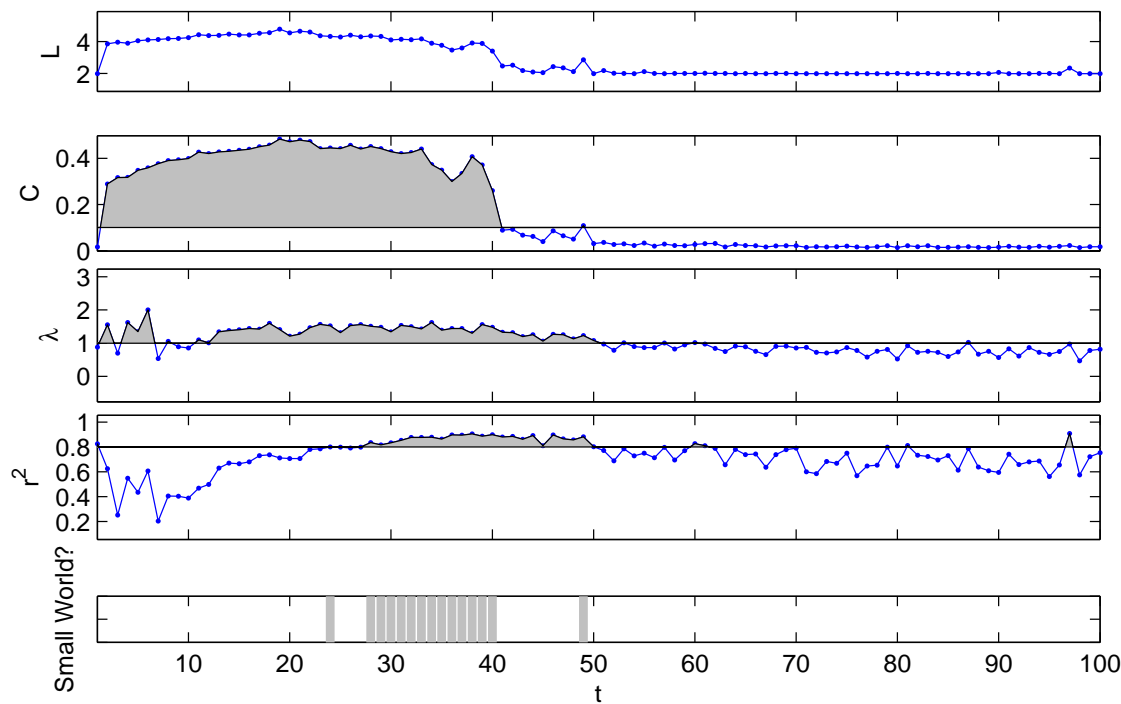}
		\caption{Small-world properties of $G=\makesw(n,m_{in},t,m)$ with respect to $t$.}
		\label{fig:swcurves}
	\end{center}
\end{figure}

It is is easy to verify that with those requirements, a random Erdös-Rényi graph with $1000$ nodes and $10000$ edges is not a small world with high probability (for instance because of the clustering coefficient). On the other hand, $G=\makesw(n,m_{in},t,m)$ verifies small-world properties for some values of $t$, as shown in Figure\ref{fig:swcurves}:
\begin{itemize}
	\item The upper curve shows the diameter $L$ (remember that we only consider the main connected component, therefore the diameter is always well defined). The diameter is always low and consistent with a small-world structure.
	\item The next curves indicates the clustering coefficient $C$. For $2\leq t\leq 40$, $C$ is very high. It drops after $40$, as the confluences converge to the nodes' degrees, meaning that most of the edges come from the highest degree nodes of the input graph. This leads to star-like structures, that explain the poor clustering coefficient.
	\item The two next curves indicates that the degree distribution may be a power-law, with a relatively high confidence, for $28\leq t\leq 50$.
	\item Lastly, the lower curve summarizes the values of $t$ that verify the small-world requirements (mainly $28\leq t\leq 40$). 
\end{itemize}

\section{Conclusion}
\label{sec:conclusion}

We proposed in this article a method to turn random graphs into Small-World graphs by the dint of random walks. This simple and intuitive method allow to set a target number of nodes and edges. The resulting graphs possess all desired properties: low diameter, low edge density with a high local clustering, and a heavy-tailed degree distribution. This method is suitable for generating random small-world graphs, but it is only a first step for answering the question: \emph{why are most of real graphs small-worlds, despite the fact that the small-world structure is very unlikely among possible graphs?}

In order to be eligible for explaining small-world effects, a small-world generator should be based on local interactions. Therefore it should be decentralized, which is not the case of Algorithm~\ref{algo:makesw}. However, there exists variations of Algorithm~\ref{algo:makesw} that can  be decentralized: for instance, if we introduce a confluence bound $s$, an algorithm where each node $u$ decide to connect with any node it can find with a mutual confluence greater than $s$ has the same behavior that Algorithm~\ref{algo:makesw} (but the number of edges $m$ is then indirectly set by the parameter $s$). Understanding the relationship between $m$ and $s$ is part of our future work.

Also note, that the random walks we used in this first algorithm may be too long: for instance, Figure~\ref{fig:swcurves} shows that a length between $28$ and $40$ is needed to achieve small-world properties for a $1000$ nodes graph, which is much larger than the expected diameter of a small-world graphs of that size. We are currently working on a way to shorten the random walks by embedding a preferential attachment scheme~\cite{albert02statistical} into our algorithm.

\bibliographystyle{abbrv}
\bibliography{proxrank}  % sigproc.bib is the name of the Bibliography in this case
\end{document}